\documentclass[aps,pra,twocolumn,superscriptaddress]{revtex4-2}

\usepackage{amsmath}
\usepackage{esint}
\usepackage{units}
\usepackage{amsbsy}
\usepackage{amssymb}
\usepackage[T1]{fontenc}
\usepackage{float}
\usepackage{graphicx}
\usepackage{dcolumn}
\usepackage{bm}
\usepackage{standalone}
\usepackage{hyperref}

\begin{document}

\title{Large Reconfigurable Quantum Circuits with SPAD Arrays and Multimode Fibers}
\author{Adrian Makowski}
\affiliation{Laboratoire Kastler Brossel, École normale supérieure (ENS) – Université Paris Sciences \& Lettres (PSL), CNRS, Sorbonne Université, Collège de France, 24 rue Lhomond, Paris 75005, France}
\affiliation{Institute of Experimental Physics, Faculty of Physics, University of Warsaw, Pasteura 5, 02-093 Warsaw, Poland}
\author{Micha\l{} D\k{a}browski}
\affiliation{Laboratoire Kastler Brossel, École normale supérieure (ENS) – Université Paris Sciences \& Lettres (PSL), CNRS, Sorbonne Université, Collège de France, 24 rue Lhomond, Paris 75005, France}
\affiliation{International Centre for Translational Eye Research, Skierniewicka 10A, 01-230 Warsaw, Poland}
\affiliation{Department of Physical Chemistry of Biological Systems, Institute of Physical Chemistry, Polish Academy of Sciences, M. Kasprzaka 44/52, 01-224 Warsaw, Poland}

\author{Ivan Michel Antolovic}
\affiliation{Advanced Quantum Architecture Laboratory (AQUA), School of Engineering, École polytechnique fédérale de Lausanne (EPFL), Rue de la Maladière, Neuchâtel CH-2002, Switzerland}

\affiliation{Pi Imaging Technology SA, 1015 Lausanne, Switzerland}

\author{Claudio Bruschini}
\affiliation{Advanced Quantum Architecture Laboratory (AQUA), School of Engineering, École polytechnique fédérale de Lausanne (EPFL), Rue de la Maladière, Neuchâtel CH-2002, Switzerland}

\author{Hugo Defienne}
\affiliation{Sorbonne Université, CNRS, Institut des NanoSciences de Paris (INSP), Paris F-75005, France}

\author{Edoardo Charbon}
\affiliation{Advanced Quantum Architecture Laboratory (AQUA), School of Engineering, École polytechnique fédérale de Lausanne (EPFL), Rue de la Maladière, Neuchâtel CH-2002, Switzerland}

\author{Radek Lapkiewicz}
\affiliation{Institute of Experimental Physics, Faculty of Physics, University of Warsaw, Pasteura 5, 02-093 Warsaw, Poland}

\author{Sylvain Gigan}
\email{sylvain.gigan@lkb.ens.fr}
\affiliation{Laboratoire Kastler Brossel, École normale supérieure (ENS) – Université Paris Sciences \& Lettres (PSL), CNRS, Sorbonne Université, Collège de France, 24 rue Lhomond, Paris 75005, France}


\begin{abstract}
Reprogrammable linear optical circuits are essential elements of photonic quantum technology implementations. Integrated optics provides a natural platform for tunable photonic circuits, but faces challenges when high dimensions and high connectivity are involved. Here, we implement high-dimensional linear transformations on spatial modes of photons using wavefront shaping together with mode mixing in a multimode fiber, and measure photon correlations using a time-tagging single-photon avalanche diode (SPAD) array. In order to prove the suitability of our approach for quantum technologies we demonstrate two-photon interferences in a tunable complex linear network –- a generalization of a Hong-Ou-Mandel interference to 22 output ports. We study the scalability of our approach by quantifying the similarity between the ideal photon correlations and the correlations obtained experimentally for various linear transformations. Our results demonstrate the potential of wavefront shaping in complex media in conjunction with SPAD arrays for implementing high-dimensional reconfigurable quantum circuits. Specifically, we achieved $(80.5 \pm 6.8)\%$ similarity for indistinguishable photon pairs and $(84.9 \pm 7.0)\%$ similarity for distinguishable photon pairs using 22 detectors and random circuits. These results emphasize the scalability and reprogrammable nature of our approach.
\end{abstract}

\maketitle

Quantum optics with indistinguishable photons have emerged as a key resource in advancing scientific research, particularly in the fields of quantum information processing\cite{nielsen_chuang_2012}, communication\cite{Mattle, doi:10.1126/science.aam9288}, and metrology\cite{giovannetti_advances_2011, hiekkamaki_photonic_2021, tenne_super-resolution_2019}, owing to their unique properties such as entanglement\cite{bouwmeester_experimental_1997, wang_multidimensional_2018, PhysRevLett.75.4337, fickler_quantum_2012}, superposition\cite{arndt_testing_2014, PhysRevLett.123.143605, hiekkamaki_high-dimensional_2021}, and non-locality\cite{PhysRevLett.47.460, PhysRevLett.81.5039}. One area of interest is the study of photonic quantum walks, which explores the behavior of quantum particles in complex environments\cite{broome_photonic_2013, 
spring_boson_2013, valencia_unscrambling_2020}. This phenomenon has potential applications in fields such as quantum algorithms\cite{PhysRevLett.102.180501}, simulations\cite{bao_very-large-scale_2023}, and metrology\cite{brandt_high-dimensional_2020, PRXQuantum.3.010202}, as well as quantum computing, communication, and sensing\cite{aslam_quantum_2023}. Several research groups have made remarkable strides in the development of quantum walk, including the first experimental realization of two-dimensional quantum walks on a lattice using single photons\cite{Schreiber}, and a quantum walk in a 21-waveguide array\cite{Peruzzo} or an application of a quantum walk to the studies of bound states between systems with different topological properties\cite{Kitagawa}. However, these experimental setups have stringent limitations regarding reprogrammability and scalability, which are crucial for scaling the system to a higher number of modes for practical implementation on near-term quantum devices\cite{madsen_quantum_2022}.

\begin{figure}[t!] 
\includegraphics[width=1\linewidth]{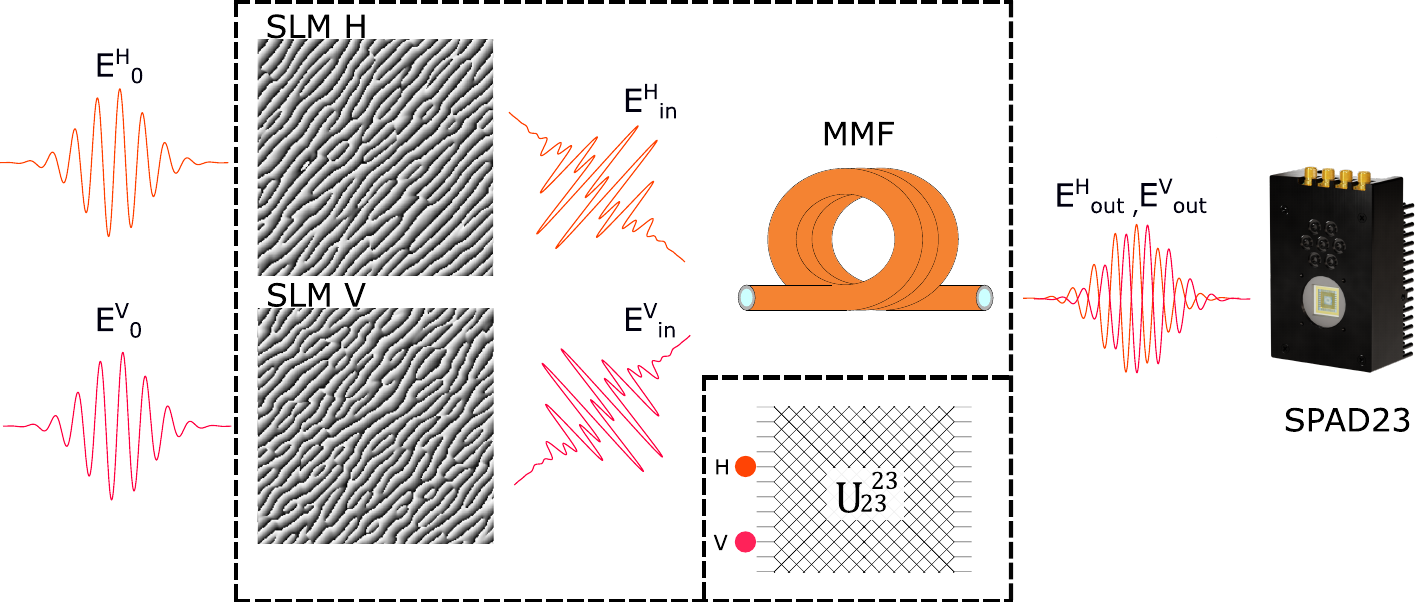} 

\caption{Reprogrammable and scalable platform for the implementation of quantum operations on a 2-photon state where wavefronts of two photons, generated using spontaneous parametric down-conversion (SPDC), are shaped using two separate parts of the spatial light modulator (SLM) before being coupled into the multimode fiber (MMF) used as a quantum states mixer\cite{Defienne,Leedumrongwatthanakun, cavailles_high-fidelity_2022}. Usage of a single-photon avalanche 23-detector array (SPAD23)\cite{SPADman} enables for subset $\mathcal{L}^{23}_{2}$ arbitrary 23-output x 2-input quantum state operations of the general $\mathcal{U}^{23}_{23}$ unitary transformation performed by MMF.}
\label{fig1a}
\end{figure}

In this letter, we present a reprogrammable and scalable platform for implementing the quantum walk of a two-photon state using a multimode fiber (MMF) as a quantum state mixer\cite{Defienne,Leedumrongwatthanakun, cavailles_high-fidelity_2022}. Our platform can generate an arbitrary N-output x 2-input quantum state operations that can be reprogrammed on demand at a 10 Hz frequency rate (see Fig. \ref{fig1a}). This provides significant advantages over existing experimental setups\cite{titchener_two-photon_2016,huisman_programmable_2015} and makes it a promising candidate for the future realizations of highly-multimode quantum walk experiments\cite{schreiber_2d_2012, crespi_anderson_2013, carolan_universal_2015, preiss_strongly_2015}. The wavefronts of the photons are shaped using a spatial light modulator (SLM), and then coupled into an MMF. The MMF is a complex medium that supports around 400 modes at wavelength $\lambda = 800$ nm and has low losses\cite{Leedumrongwatthanakun} which  are the essential features for performing multidimensional unitary operations on the single photons efficiently \cite{li_compressively_2021, Defienne,defienne_arbitrary_2020, morales-delgado_two-photon_2015}.

Previous implementations faced a significant limitation in achieving scalability due to the integration of detection technology\cite{Leedumrongwatthanakun}.This required a large number of separate avalanche photodiodes, rendering the solution of problems like boson sampling \cite{brod_photonic_2019}, very impractical. Several experiments have employed single-outcome projective measurements for sequential analysis of the output state \cite{goel2022inversedesign, valencia_unscrambling_2020}. However, these approaches suffer from inherent limitations, including substantial losses (as only two outputs can be detected simultaneously out of all possibilities) and time-intensive procedures (since detecting $i$ output modes demands $i^K$ measurements, with $K$ representing the number of photons involved). Consequently, these methods become impractical for large-scale systems. 

To overcome this issue, here we use a 23-single photon avalanche detector (SPAD23)\cite{SPADman} array to measure the number of counts and coincidences between photons at the linear network output.  The ability to modify the phase pattern on the SLM allows the MMF to performs a specific linear operation on a two-photon state. This operation can be easily and reliably adjusted on demand.

Degenerate photon pairs at 810nm are produced by type-II spontaneous parametric down conversion (SPDC) process in a ppKTP crystal pumped by a 405 nm continuous wave laser. To split the photons with orthogonal polarizations, we use a polarisation beam splitter (PBS). Our SPDC setup allows us to adjust the time delay between the photons, and then observe and control Hong-Ou-Mandel (HOM) interference\cite{PhysRevLett.59.2044} by changing the photons' distinguishability. The measured HOM visibility of photons from our SPDC source is approx. 95\%. See \textit{Supplemental Material} for more details on the source.

In our experiment, we utilized a detector array consisting of Single-Photon Avalanche Diodes (SPADs) using CMOS technology\cite{Lubin:19, ghezzi_multispectral_2021}, specifically the SPAD23 model from Pi Imaging Technology\cite{SPADman}. This detector offers a sub-ns temporal resolution (120 ps jitter FWHM and 20 ps for least-significant bit when using time-to-digital converters as time-taggers), exhibits low dark noise (less than 100 counts per second at 20$^o$C), has a high pixel fill factor (80\%), and a "dead time" of approximately 50 ns\cite{Antolovic:18}. However, like all SPAD arrays SPAD23 is prone to cross-talk, which occurs when a photon detected by one of the array's detectors is simultaneously counted by a neighboring detector\cite{bruschini_single-photon_2019, Lubin:19}. While the probability of cross-talk is low (approximately 0.1\%), it affects the number of coincidences measured in our experiment but not the number of single photon counts\cite{Lubin:19}. To account for cross-talk, we employed a calibration procedure, which is detailed in the \textit{Supplemental Material}.

\begin{figure}[b!] 
\includegraphics[width=1\linewidth]{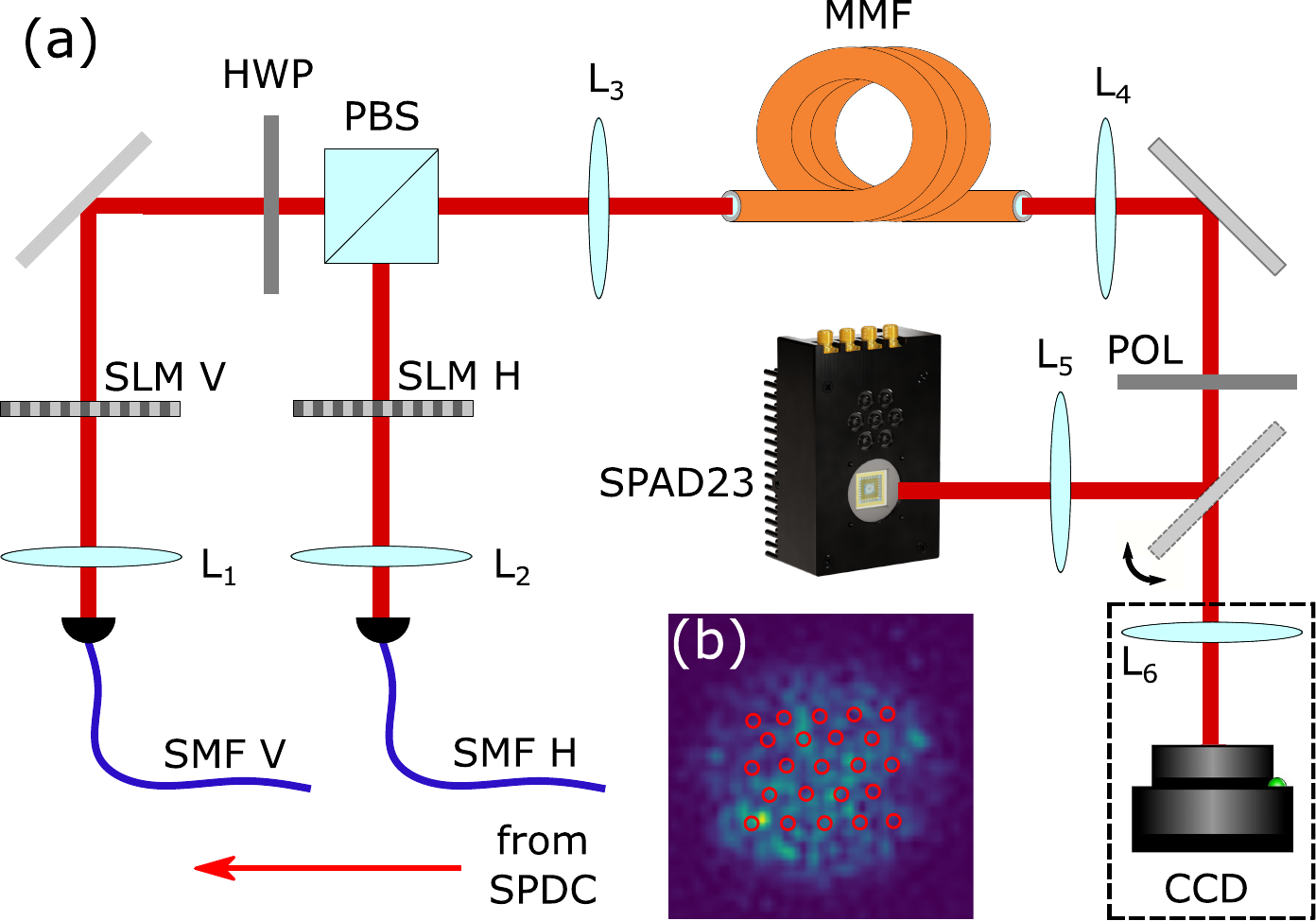} 

\caption{(a) Schematic of the experimental setup. A photon pair of two orthogonal polarizations (H and V) is passed through single-mode fibers. The photons are collimated, and their wavelengths are shaped by the spatial light modulator (SLM). The shaped photons are then coupled into a multimode fiber (MMF). The MMF output is imaged on the SPAD23 or CCD by changing the flip-mirror position to measure the number of counts and coincidences. By knowing the MMF's transmission matrix, we can select a phase pattern on the SLM to perform an arbitrary operation. (b) Speckle image of the light coming out of the fiber imaged on the camera, and the localization of the SPAD23 detectors.}
\label{fig1b}
\end{figure}

Figure \ref{fig1b} depicts the experimental setup used in our study. Two separate parts of the spatial light modulator (SLM), labeled as H and V for orthogonal light polarizations, were illuminated with two photons created in our SPDC platform. The SLM shaped the wavefronts of the photons, which were then focused on the MMF of 50$\mu m$ core. The MMF with SLM induces a specific quantum operation on a 2-photon state, and modifying the phase pattern on the SLM allows for easy adjustment of this operation. The resulting speckle image of the light emerging from the MMF was either imaged on the SPAD23 or the CCD camera after passing through a polarizer to choose just one particular polarization (for which the TM was measured).  To calibrate the relative position of the MMF and SPAD23, we used a CCD camera, as depicted in Fig. \ref{fig1b}(b). It shows the speckle image captured by the CCD camera, with the positions of the SPAD23 detectors marked with red circles. The CCD camera was used just for the calibration purposes (see \textit{Supplemental Material} for details). The magnification was choosen to map one speckle mode of the MMF into one single SPAD detector. The photon arrival time was measured using SPAD23 detector and then used to calculate the number of counts $n_i$ for each detector $i$ and coincidences $C_{ij}=\langle n_i n_j \rangle_t$ for each pair $(i, j)$.

\begin{figure}[t!] 
\includegraphics[width=1\linewidth]{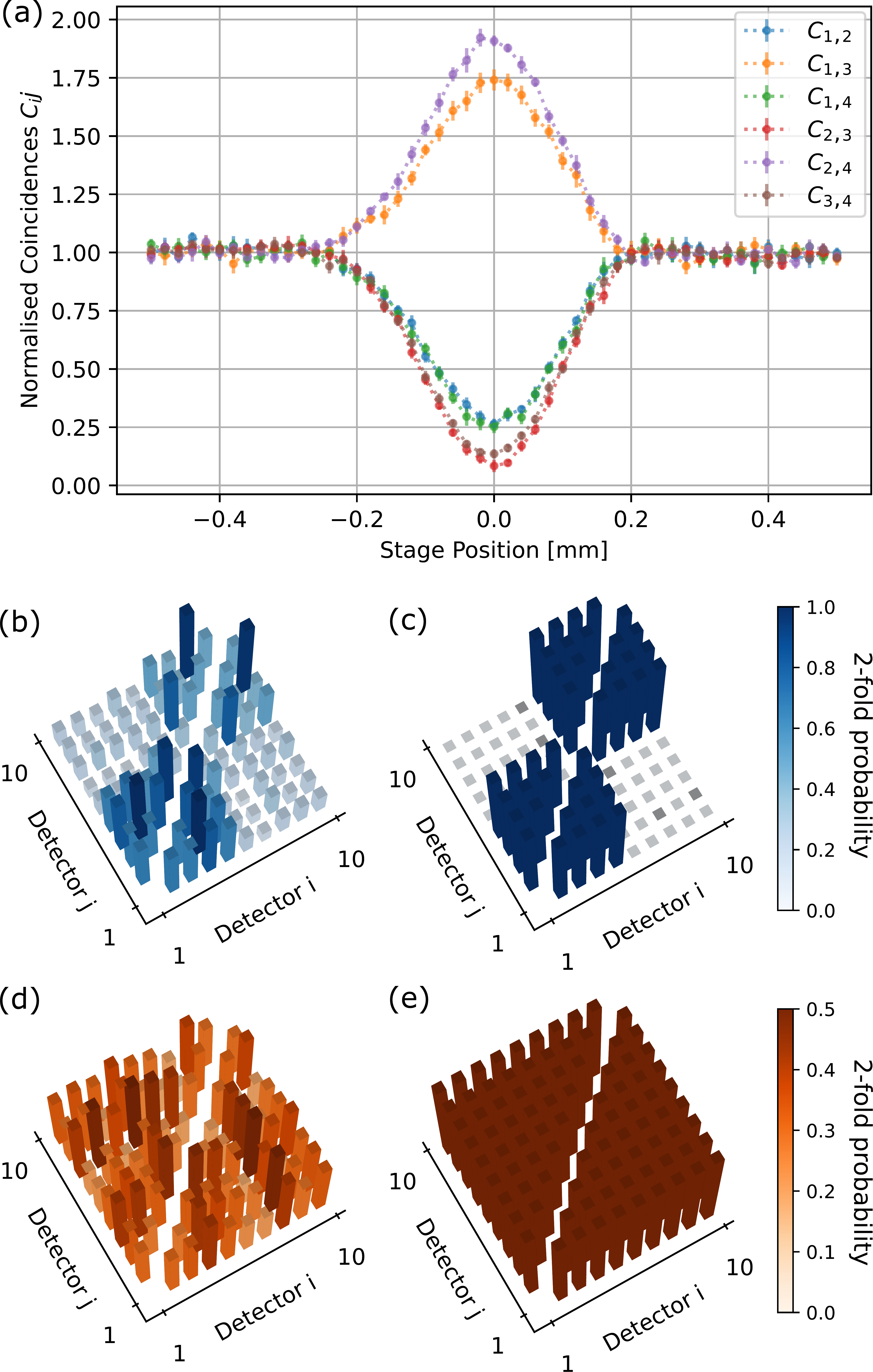} 

\caption{(a) Hong-Ou-Mandel (HOM) interference scan for a 4x4 operation, displaying the number of coincidences as a function of the relative delay between the two photons. (b)-(e) Example results of a 10x10 operation on the 2-photon state. (b) and (c) show the experimental and theoretical coincidence counts for indistinguishable photon pairs, respectively. (d) and (e) show the corresponding experimental and theoretical coincidence counts for distinguishable photon pairs, acquired over a 100-second measurement period. The SPAD23 detectors used in the experiment are marked in yellow in Fig. \ref{fig3}.}
\label{fig2}
\end{figure}

Our experimental platform establishes a connection between input modes displayed on the SLM and corresponding output modes measured using 23 detectors described by unitary operator $\mathcal{U}_{23}^{23}$, using the well-established technique of a transmission matrix (TM) measurement of the optical system\cite{Popoff}. The measured TM is stable for a few days in normal laboratory conditions. We measure the TM in a Fourier mode basis by displaying phase ramps on the SLM with a varying inclination and orientation\cite{Loterie:15}. This allows us to scan the different spatial positions at the entrance of the MMF and as a result to address particular output modes of the MMF after TM calibration. If the addressed mode is not an eigenvector of the TM, the light becomes scrambled as it propagates through the MMF. The electric field amplitude at SPAD23 is linearly dependent on the electric field at the SLM and can be represented as:

\begin{equation}
E^{(k)}_{out}=T^{(1)}_k E^{(k)}_{in}\text{, for }k=H,V,
\end{equation}

where $E^{(k)}_{out}$ is the electric field at SPAD23, $T^{(1)}_k$ is a one-photon transmission matrix for SLM part $k$, and $E^{(k)}_{in}$ is the electric field on the SLM part $k$, corresponding to the SLM part shaping the polarisation $k=H, V$. During the TM measurement, the SPAD acquires the number of photon counts per 10 ns time-window for each SLM mode, and in order to obtain the electric field on SPAD23 we perform phase-stepping interferometry\cite{Popoff}. We perform this operation separately for both light polarization (SLM parts) and then calculate the transmission matrix for the two-photon state\cite{Defienne}:

\begin{equation}
T^{(1)}_H, T^{(1)}_V \rightarrow T^{(2)}.
\end{equation}

With this information, we can readily calculate the SLM pattern that gives us the required quantum operation on the two-photon state $\mathcal{L} \in \mathbb{M}_{2\mathsf{x}N}$, where $N$ is the number of detectors. The computation of SLM pattern for a given $\mathcal{L}$ takes only a few seconds. The electric field on SLM can be calculated using the complex conjugate of the transmission matrix for the two-photon state:

\begin{equation}
[E^{(H)}_{in}, E^{(V)}_{in} ]=T^{\dagger(2)}\mathcal{L}.
\end{equation}

Knowing the TM of the MMF, one can modify the phase pattern at the speed of 10Hz on the SLM and obtain different N-output x 2-input quantum linear network operations $\mathcal{L}_{2}^{N}$.

As an example of an all-to-all operator, we emulate a 10x10 Sylvester operation\cite{Viggianiello_2018} on the two-photon state generated by the experimental platform described above. We measured the TM of the MMF in our setup to calculate the appropriate SLM pattern for performing the 10-dimensional Sylvester operation:

\begin{equation}
\mathcal{L}_S = \left\{ \begin{array}{rcl}
1 & \mbox{for}
& k=V \\ (-1)^i & \mbox{for} & k=H, 
\end{array}\right.
\end{equation}
where $i$ denotes the detector index. We measured the number of coincidence counts for distinguishable and indistinguishable photon pairs.

\begin{figure}[t!] 
\includegraphics[width=0.97\linewidth]{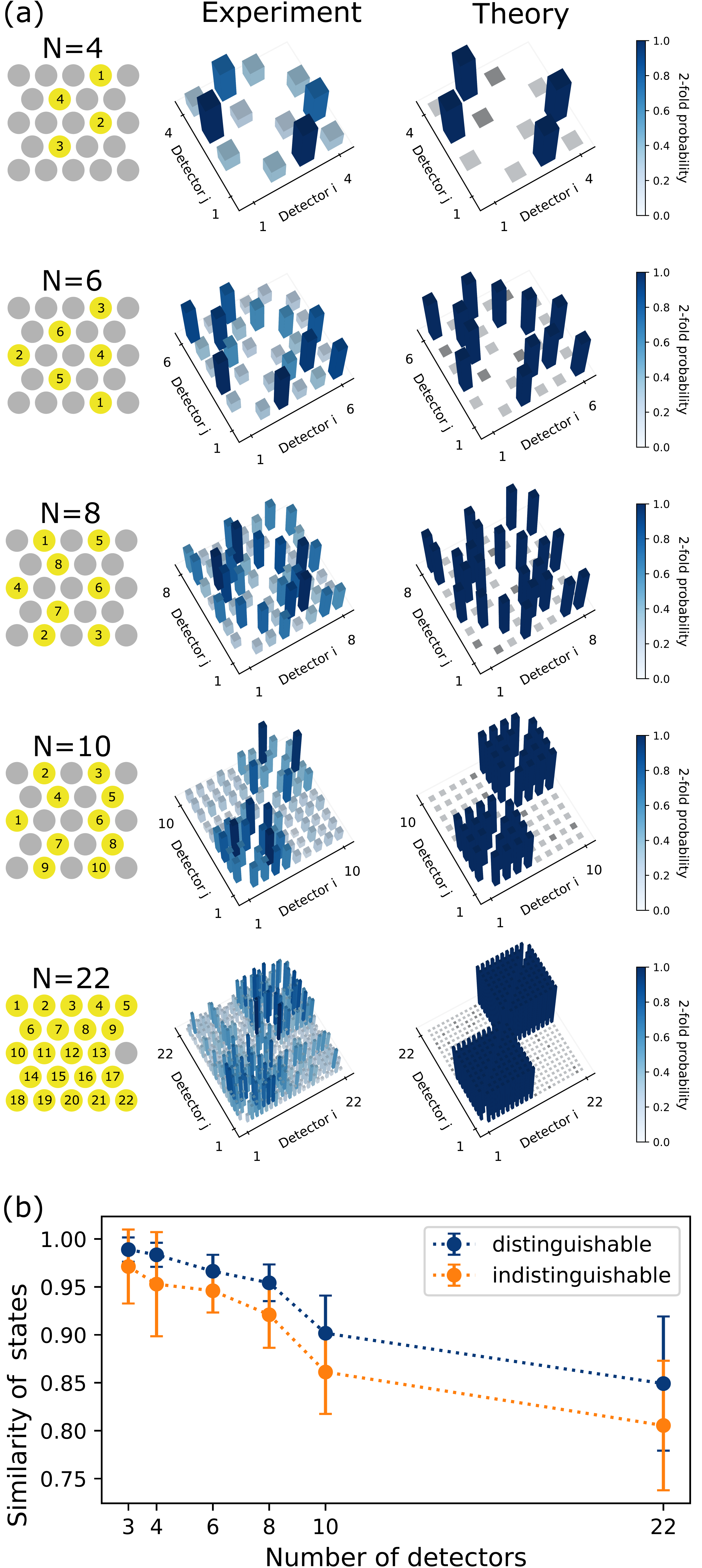} 

\caption{(a) Sylvester transformation for different numbers of detectors. The first column exhibits the SPAD23 detector array used in the experiment, with the detectors used for the measurement with a given number of detectors marked in yellow. The second column shows the experimental coincidence counts for various numbers of detectors and indistinguishable photon pairs. The third column shows the theoretical coincidence counts for the corresponding number of detectors. (b) A random operation similarity trend for indistinguishable and distinguishable photon pairs.}
\label{fig3}
\end{figure}

The results of this experiment are presented in Fig. \ref{fig2}. Figure \ref{fig2}(a) shows a HOM interference scan for a 4-dimensional Sylvester operator, which corresponds to the number of coincidences as a function of the relative delay between the two photons. The HOM dip in the scan indicates the presence of interference between the two photons, which is essential for quantum operations with indistinguishable particles\cite{PhysRevLett.59.2044}. The HOM visibility $V$ for different coincidence distributions $C_{i,j}$, ranging from $V=0.74$ to $V=0.92$ deviates from ideality (the 95\% indistinguishably of the source) mainly because of cross-talk between different SPAD23 detectors (see \textit{Supplemental Material} for details) as well as photons spectral dispersion when propagating through the MMF and non-perfect fidelity of linear network operator.

Figures \ref{fig2}(b) and \ref{fig2}(d) show the experimental coincidence counts for indistinguishable and distinguishable photon pairs, respectively over 10 output modes. These counts were acquired for 100 sec by measuring the number of coincidences between the SPAD23 detectors when the photons were either indistinguishable or distinguishable. The theoretical coincidence counts for indistinguishable and distinguishable photon pairs\cite{Vlasov_2017, scheel2004permanents} are presented in Fig. \ref{fig2}(c) and \ref{fig2}(e), respectively.

Comparing the experimental results presented in Figs. \ref{fig2}(b) and \ref{fig2}(d) with the theoretical predictions in Figs. \ref{fig2}(c) and \ref{fig2}(e), we can see that the presented results agree well with the theoretical predictions\cite{Vlasov_2017, scheel2004permanents}.
The experimental results also demonstrate that the experimental platform can successfully generate two-photon states for quantum operations, as well as measure and characterize their properties through coincidence counting.

We finally conducted an investigation into the scalability and reprogrammable nature of our experimental platform by performing random matrix quantum operations with varying numbers of detectors. Figure \ref{fig3}(a) presents coincidence distributions, with the first column showcasing the theoretical coincidence counts for indistinguishable photon pairs utilizing the Sylvester operation\cite{Viggianiello_2018} across a range of detectors. Meanwhile, the second column demonstrates the corresponding experimental coincidence counts, which we determined by measuring the number of coincidences between the SPAD detectors. Because of a high number of dark counts in one of the detectors, we excluded it from the experiments.

Figure \ref{fig3}(b) shows the difference between the similarity trend for distinguishable and indistinguishable photon pairs $\langle\mathcal{S}_{ET}\rangle_{\mathcal{L}_R}$ when performing 100 random $\mathcal{L}_R$ operations (random complex numbers from a uniform distribution) for different number of SPAD detectors. The similarity $\mathcal{S}_{ET}$ of two coincidence distributions $C^{(E)}_{i,j}$ and $C^{(T)}_{i,j}$ (corresponding to particular $\mathcal{L}_R$ operator), representing a generalized fidelity for 2-fold coincidences\cite{Peruzzo}, is defined as:
\begin{equation}
    \mathcal{S}_{ET}=\frac{\left(\sum_{i,j}\sqrt{C^{(E)}_{i,j}C^{(T)}_{i,j}}\right)^2}{\sum_{i,j}C^{(E)}_{i,j}\sum_{i,j}C^{(T)}_{i,j}}.
\end{equation}

In other words, similarity quantifies the extent to which the experimental results align with theoretical predictions, with higher values indicating stronger agreement.  We see that the similarity decreases as we increase the number of detectors from 4 to 22, from $98.3 \pm 1.23\%$ ($95.3 \pm 5.5$\%) to $84.9 \pm 7.0\%$ ($80.5 \pm 6.8\%$) for distinguishable (resp. indistinguishable) photons. The similarity is higher for distinguishable pairs because of more stringent conditions for 2-photon interference (distinguishable photons are not affected by phase errors in quantum interference). For the same reason, the similarity for indistinguishable pairs of photons is reduced due to the limited accuracy of photon wavefront-shaping via SLM. Also, for a larger number of detectors, the similarity decreases because the coincidence distribution becomes more noisy as can be seen in Fig. \ref{fig3}.

To conclude, we present an approach to implementing high-dimensional reconfigurable quantum circuits using wavefront shaping and mode mixing in the MMF with a SPAD camera as a detector. It offers advantages in terms of scalability and flexibility compared to other approaches\cite{huisman_programmable_2015, wolterink_programmable_2016, PhysRevLett.118.093902}. We demonstrate the feasibility of the presented approach by implementing a complex linear network for two-photon interference. We measure the two-photon correlations between all the output pairs using a time-tagging SPAD array\cite{bruschini_single-photon_2019, slenders_confocal-based_2021}, thus advancing towards scalable detection schemes beyond previously proposed solutions\cite{Defienne, Leedumrongwatthanakun}. The similarity between the ideal photon correlations and the correlations has been obtained experimentally for up to 22 output modes for various randomly chosen linear transformations for both indistinguishable and distinguishable photon pairs. The current limitation of the setup is the number of available photons preventing us from studying high dimensional linear networks applied to multi-photon states  $\mathcal{L}$\cite{wang_boson_2019, doi:10.1126/science.aar7053}, as well as detector cross-talk\cite{Lubin:19} that affect coincidences between detectors located close to each other thus reducing the measured similarity of the states.

Future work in this area could focus on further optimizing the wavefront shaping and mode mixing techniques -- along with sources delivering more than two photons enabling the achieving of even higher dimensional transformations\cite{wang_boson_2019}. Additionally, the use of more advanced detectors, either superconducting nanowire single-photon detectors with near-unity quantum efficiency\cite{Reddy:20} or SPAD cameras with more pixels\cite{s18114016,bruschini_single-photon_2019} could further improve the measurement capabilities. The scalability and programmable nature of the presented approach make it promising for applications in quantum information processing, such as quantum communication\cite{Mattle, doi:10.1126/science.aam9288} and quantum computing\cite{schreiber_2d_2012, crespi_anderson_2013, carolan_universal_2015, preiss_strongly_2015}, especially in the perspective of using more detectors to test different boson sampling protocols which can overcome the capabilities of existing classical information processing schemes\cite{madsen_quantum_2022, wang_boson_2019, lund_quantum_2017, harrow_quantum_2017}.

\textit{Acknowledgments} We would like to thank Saroch Leedumrongwatthanakun for fruitful discussions and initial guidance with the setup. S.G. acknowledges funding from the European Research Council ERC Consolidator Grant (SMARTIES-724473). H.D. acknowledges funding from the European Research Council ERC Starting Grant (SQIMIC-101039375). A.M acknowledge Scholarship of French Government - Ph.D. Cotutelle/Codirection and "IV.4.1. A Complex Programme of Support For UW PhD Students". A.M. and R.L acknowledge the support by the Foundation for Polish Science under the FIRST TEAM project
‘Spatiotemporal photon correlation measurements for quantum metrology and super-resolution microscopy’ co-financed by the European Union under the European Regional Development Fund (POIR.04.04.00-00-3004/17-00). The International Centre for Translational Eye Research (MAB/2019/12) project is carried out within the International Research Agendas Program of the Foundation for Polish Science, co-financed by the European Union under the European Regional Development Fund.

\nocite{*}
\bibliographystyle{apsrev4-1}
\bibliography{apssamp} 
\end{document}


\title{Large Reconfigurable Quantum Circuits with SPAD Arrays and Multimode Fibers -- Supplemental Material}

\author{Adrian Makowski}
\affiliation{Laboratoire Kastler Brossel, École normale supérieure (ENS) – Université Paris Sciences \& Lettres (PSL), CNRS, Sorbonne Université, Collège de France, 24 rue Lhomond, Paris 75005, France}
 \affiliation{Institute of Experimental Physics, Faculty of Physics, University of Warsaw, Pasteura 5, 02-093 Warsaw, Poland}
  
\author{Micha\l{} D\k{a}browski}
\affiliation{Laboratoire Kastler Brossel, École normale supérieure (ENS) – Université Paris Sciences \& Lettres (PSL), CNRS, Sorbonne Université, Collège de France, 24 rue Lhomond, Paris 75005, France}
\affiliation{International Centre for Translational Eye Research, Skierniewicka 10A, 01-230 Warsaw, Poland}
\affiliation{Department of Physical Chemistry of Biological Systems, Institute of Physical Chemistry, Polish Academy of Sciences, M. Kasprzaka 44/52, 01-224 Warsaw, Poland}

\author{Ivan Michel Antolovic}
\affiliation{Advanced Quantum Architecture Laboratory (AQUA), School of Engineering, École polytechnique fédérale de Lausanne (EPFL), Rue de la Maladière, Neuchâtel CH-2002, Switzerland}
\affiliation{Pi Imaging Technology SA, 1015 Lausanne, Switzerland}

\author{Claudio Bruschini}
\affiliation{Advanced Quantum Architecture Laboratory (AQUA), School of Engineering, École polytechnique fédérale de Lausanne (EPFL), Rue de la Maladière, Neuchâtel CH-2002, Switzerland}

\author{Hugo Defienne}
\affiliation{Sorbonne Université, CNRS, Institut des NanoSciences de Paris (INSP), Paris F-75005, France}

\author{Edoardo Charbon}
\affiliation{Advanced Quantum Architecture Laboratory (AQUA), School of Engineering, École polytechnique fédérale de Lausanne (EPFL), Rue de la Maladière, Neuchâtel CH-2002, Switzerland}

\author{Radek Lapkiewicz}
\affiliation{Institute of Experimental Physics, Faculty of Physics, University of Warsaw, Pasteura 5, 02-093 Warsaw, Poland}

\author{Sylvain Gigan}
\email{sylvain.gigan@lkb.ens.fr}
\affiliation{Laboratoire Kastler Brossel, École normale supérieure (ENS) – Université Paris Sciences \& Lettres (PSL), CNRS, Sorbonne Université, Collège de France, 24 rue Lhomond, Paris 75005, France}

\begin{abstract}
\textbf{Abstract:} This Supplementary Material consist of three sections extending the content of the main manuscript. Section SI presents the detailed description of the experimental setup used for operation on photon pairs. Section SII shows the calibration of the position of SPAD23 array in relative to the multimode fibre using CCD camera. Section SIII describes the procedure for accidential coincidences and cross-talk subtraction from the measured data. 
\end{abstract}

\maketitle

\section*{SI. Experimental setup details}

\begin{figure*}[t!]
  \centering
  \includegraphics[width=1\linewidth]{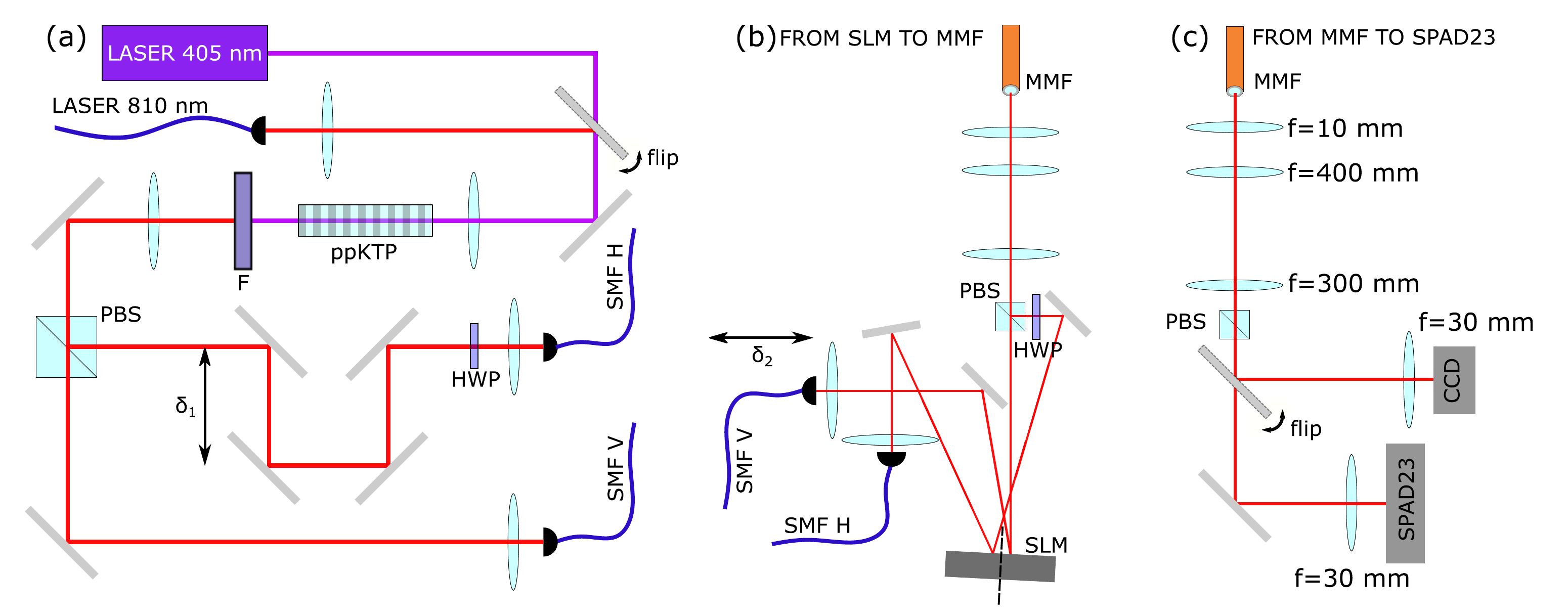}
\caption{Full scheme of the experimental setup. (a) Photon-pairs generating module consist of cw-laser which pumps ppKTP nonlinear crystal. In this process, two orthogonaly-polarized photons are emitted and then collected via separate single-mode fibers (SMF). Half-wave plate anong with delay line $\delta_1$ were used to obtain the same polarization and time-overlapping of both photons. Instead of blue-light emitting laser, the NIR source can be used for calibration purposes. F - narrow-band spectral filter, PBS - polarizing beam splitter (b) SLM-based photon shaping module. After adjusting the relative distance $\delta_2$ of both photons from the spatial light modulator (SLM), they are shaped by distinct parts of the SLM and subsequently impinging into multimode fibre (MMF) placed in the Fourier plane relative to the SLM. As the SLM is working for one particular polarization of light, the second photon is made polarization-orthogonal after SLM reflection. (c) SPAD23 or CCD camera detection module. Using flip mirror we can choose SPAD23 or CCD as a detector. Because we control only one polarization during TM measurement (and the detectors are polarization-insensitive), additional PBS is placed in front of the chosen detector to realize linear network operator $\mathcal{L}$ for one particular output polarization of light.}
\label{Fig_S_Set}
\end{figure*}

The detailed scheme of our photon pairs source is presented in Fig. \ref{Fig_S_Set}(a). A two-photon state is generated using the type-II spontaneous parametric down-conversion (SPDC) process in a 10-mm periodically poled potassium titanyl phosphate crystal (ppKTP). The crystal is pumped by a 405 nm single-mode continuous-wave laser (DLproHP, Toptica) in a single spatial mode configuration. In the process two orthogonaly-polarized photons with a wavelength of 810 nm defined by a bandpass filter with a width of 1 nm are generated. To split the photons with orthogonal polarization, we use a polarisation beam splitter (PBS). By adjusting the time delay between the photons ($\delta_1$ in Fig. \ref{Fig_S_Set}a and $\delta_2$ in Fig. \ref{Fig_S_Set}b), we can control the Hong-Ou-Mandel interference\cite{PhysRevLett.59.2044} and measure its properties, as we change the distinguishability of the photons. Using a half-wave plate we change the polarization of one of the photon to make both of them indistinguishable when shaping by the spatial light modulator (SLM). The measured Hong-Ou-Mandel visibility of photons from our SPDC source is approx. 95\%. Moreover, we can exchange photon source to a classical laser light with a wavelength of 810 nm. Thus we can perform calibration procedures such as cross-talk level measurements (described in section SIII) and calibration of SPAD23 and MMF relative positions much faster (section SII).

Figure \ref{Fig_S_Set}(b) depicts the setup module used for the photons wavefront shaping. Two separate parts of the SLM (Hamamatsu X10468-02, resolution 800x600 pixels, 20 $\mu$m pixel size) labeled as H and V, were illuminated with two previously splitted photons created by SPDC module (Fig. \ref{Fig_S_Set}(a)). The SLM is shaping the wavefronts of the photons, which are next merged on PBS, and then directed into MMF (Thorlabs, GIF50C 50$\pm$2.5$\mu m$ core diameter, 55.3$\pm$0.1 cm length, NA = 0.2) placed in the conjugate plane relative to the SLM. The MMF along with SLM induces a specific linear network quantum operation $\mathcal{L}$ on a 2-photon state \cite{Defienne,Leedumrongwatthanakun, cavailles_high-fidelity_2022}, and modifying the phase pattern $s_i$ on the SLM allows for easy change of this operator properties. The resulting speckle patterns from the MMF were either imaged on the SPAD23 (Pi Imaging Technology) or the CCD camera (FLIR Point Grey: CMLN-13S2M-CS) after passing through a polarizing beam splitter (to choose just one particular polarization for which the TM was measured). The CCD camera (resolution 1296 x 964 pixels, pixel size of 3.75 $\mu$m) has just been used to calibrate the relative position of the MMF and SPAD23, as described in section SII. We used 4-fold magnification to adjust the size of one speckle mode of the MMF into single SPAD detector. The photon arrival times are measured using SPAD23 detector and then we use Matlab scripts to calculate the number of counts $n_i$ for each detector $i$ and coincidences $C_{ij}=\langle n_i n_j \rangle_t$ for each pair $(i, j)$ per second in the post-processing.

As a detector, we utilized  SPAD23 model from Pi Imaging Technology\cite{SPADman}, a detector array consisting of Single-Photon Avalanche Diodes (SPADs) using CMOS technology\cite{Lubin:19, ghezzi_multispectral_2021}. This detector offers a sub-ns temporal resolution (120 ps jitter level), exhibits low dark noise (less than 100 counts per second at 20$^o$C), and a "dead time" of approximately 50 ns\cite{Antolovic:18}. The single detectors are arranged in a hexagonal grid pattern with a minimum distance of 23 $\mu$m between adjacent detectors. The radius of the single detector active area (native 24\% fill factor increased to 80\% using array of microlenses) was c.a. 5.85 $\mu$m. However, like all SPAD arrays SPAD23 is prone to cross-talk, which occurs when a photon detected by one of the array's detectors is simultaneously counted by a neighboring detector\cite{bruschini_single-photon_2019, Lubin:19}. While the probability of cross-talk is low (approximately 0.1\%), it affects the number of coincidences measured in our experiment but not the number of photon counts\cite{Lubin:19}. To account for cross-talk, we employed a calibration procedure described in details in section SIII.

\section*{SII. Adjustment of relative position between MMF and SPAD23} \label{Adjustment_position}

\begin{figure*}[t!]
  \centering
  \includegraphics[width=1\linewidth]{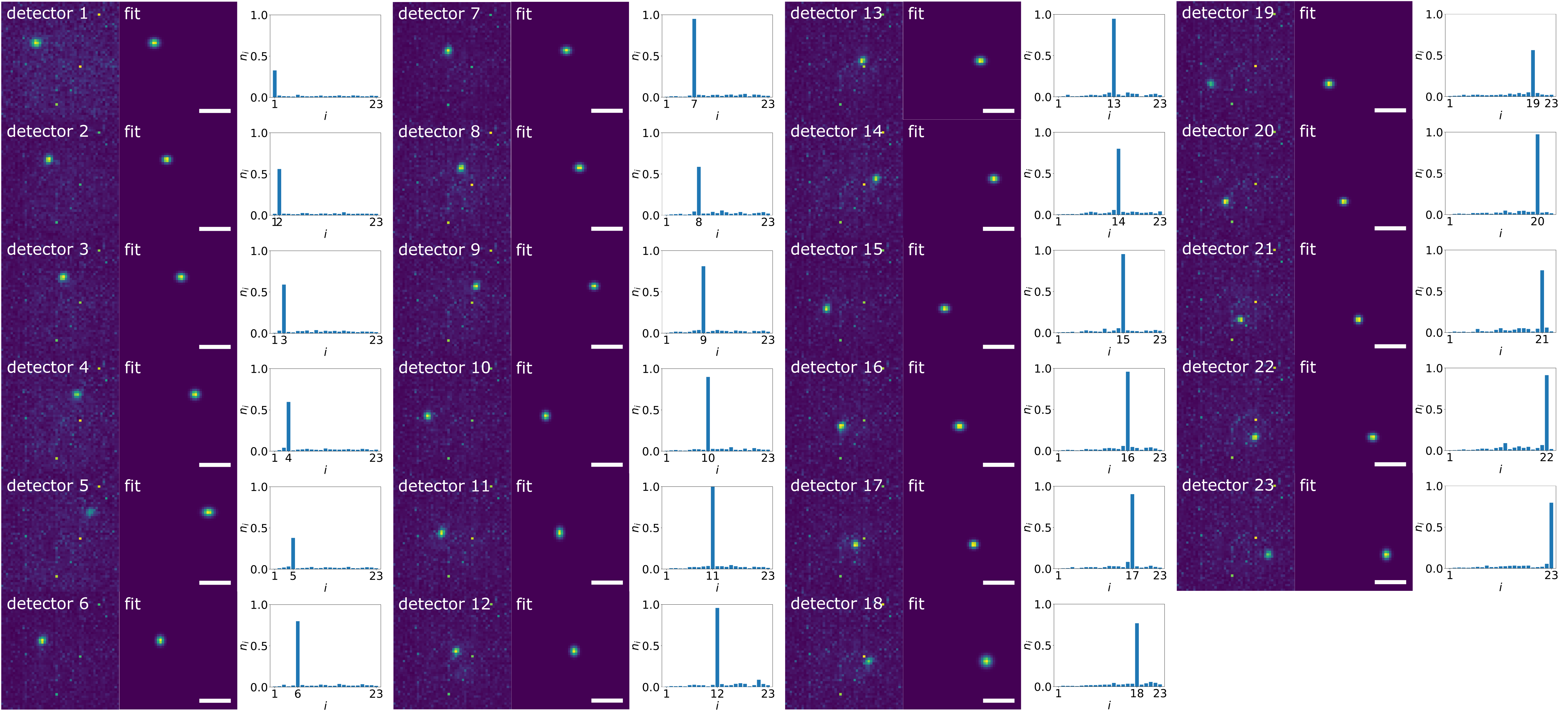}
\caption{Calibration of relative position between SPAD23 and multimode fiber (MMF). CCD images of focused light on camera position corresponding to specific SPAD23 detector, along with 2D Gaussian functions fitted to each image and number of counts collected by each SPAD23 detector are presented, respectively. As shown, we obtain good quality of focusing on any of the detectors. All CCD camera images as well as histograms of SPAD23 detector counts are normalize to the maximum value registered for one particular detector. This ensures reliable measurements using SPAD23 detectors. All camera images are in the same scale, white bar equals $50 \mu$m.}
\label{det_vs_SPAD}
\end{figure*}

\begin{figure}[b!]
  \centering
  \includegraphics[width=1\linewidth]{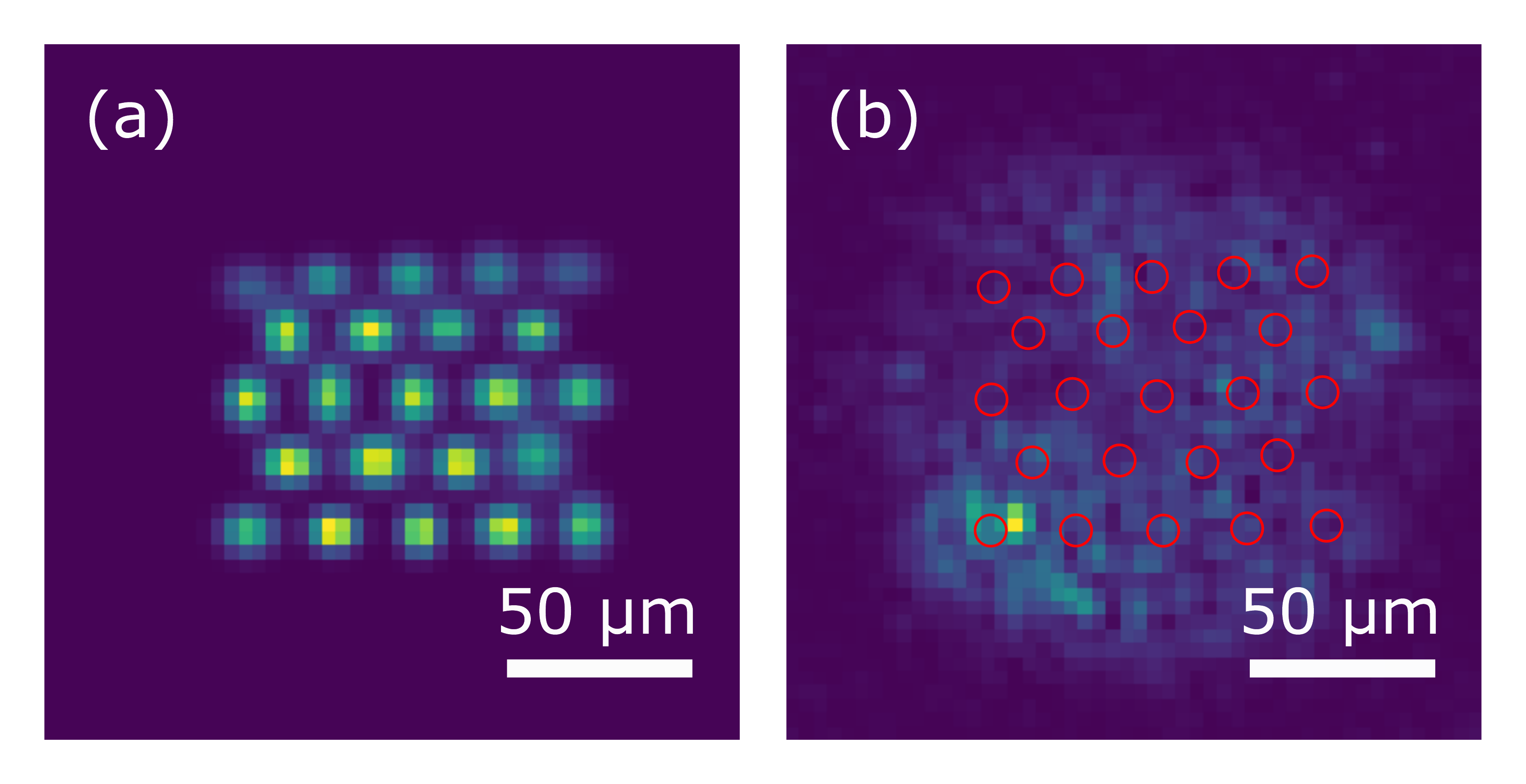}
\caption{(a) The sum of all Gaussian fits from Fig. \ref{det_vs_SPAD} shows the regular structure of SPAD23 detectors placed on a hexagonal grid. (b) CCD camera picture of a typical speckle pattern from MMF along with marked positions (in red) of the detectors extracted from (a). Each red circle has a diameter of two standard deviations of a corresponding Gaussian fit.}
\label{Fig_S_all_foc_specl}
\end{figure}

To ensure an accurate alignment of the SPAD23 detectors in relative to the multimode fiber (MMF), we developed a calibration procedure decided below. It involves using a classical coherent light source, a CCD camera, and a spatial light modulator (SLM) to set a phase pattern corresponding to focusing light on specific SPAD23 detector. First, we measured a transmission matrix (TM) of the MMF fiber with the SPAD23 array and a laser as the classical source of coherent light with Poissonian statistics. After TM measurement we have changed the output detector from SPAD23 to a CCD camera. Next, based on TM calibration we display on the SLM the phase patterns $s_i$ that focus light scrambled after propagation through MMF on a particular SPAD23 detector. We captured the images of the focused light as well as the random speckle patterns corresponding to the random phase patterns displayed on the SLM, with the CCD camera. To ensure our calibration procedure works correctly, we then replaced CCD by SPAD23 and repeat the same procedure, obtaining histograms of photon counts on each SPAD23 detector for a particular set of SLM patterns $s_i$. To each CCD camera image of of the focused light spot, we fitted a 2D-Gaussian function to determine its position accurately. Then we localized the positions of the fitted Gaussian functions within the region where random speckle patterns can be registered. Figure \ref{det_vs_SPAD} displays CCD images of focused light on specific SPAD23 detector along with a corresponding 2D Gaussian function. Additionally, the sum of all fits is also shown is Fig. \ref{Fig_S_all_foc_specl} along with a random speckle pattern with marked SPAD23 detector positions recovered from such fitting procedure. This calibration process ensures the accurate alignment of the SPAD23 detectors with the MMF fiber, enabling reliable measurements in the experiments presented in the main manuscript.

\section*{SIII. Accidental coincidences and cross-talks correction procedure}

SPAD arrays' compactness, high quantum efficiencies, and time resolution allow them to become widely used tools in many experimental applications and technology. However, one of the inconveniences of using them is a phenomenon called cross-talk. Cross-talk occurs when a photon coming to one of the array's detectors is counted by another neighboring detector simultaneously. It is a rare phenomenon (around 0.1\% probability for SPAD23 used in our experiment), therefore it does not affect the photon count rate. Nevertheless, it strongly influences the number of coincidences, which is critical in quantum applications.

Another phenomenon that causes fake coincidence detection is so-called accidental photons. Two photons arrive at two different detectors simultaneously. However, they weren't created together in the SPDC source, and they are distinguishable. Photons just accidentally appeared on both detectors at the same time. The likelihood of this phenomenon is proportional to the product of the number of photons per second on both detectors.
 \begin{figure}[t!]
  \centering
  \includegraphics[width=1\linewidth]{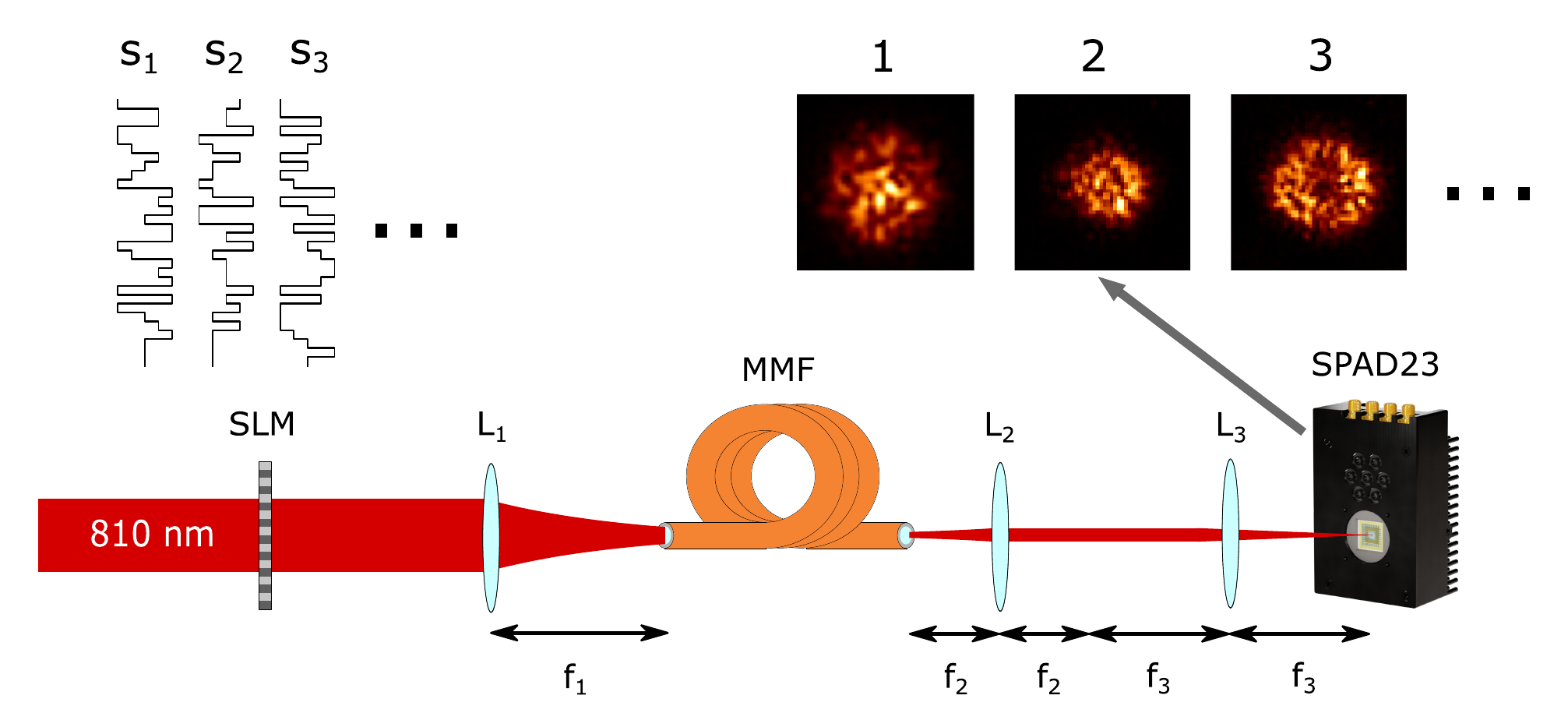}
\caption{Configuration of the experiment used for measuring accidental coincidences $\mathbf{\alpha}$ and cross-talks counts $\mathbf{\beta}$ matrices with a classical light laser source. There are shown several example images registered by the CCD camera placed in a position of SPAD23, when displaying different random patterns $s_i$ on the SLM. SPAD23 detectors registered random number of counts for each particular $s_i$ SLM pattern which in the end enable us sampling the SPAD23 detectors responses on the whole distribution of available photon counts.}
\label{Cross-talk_setup}
\end{figure}
To estimate an amount of accidental counts and cross-talk we perform an experiment presented in Fig. \ref{Cross-talk_setup}. We estimate these values with classical light generated by the laser. We shape the wavefront of the laser beam by displaying a random phase pattern $s_p$ on the SLM.  The light is coupled into the multi-mode fiber, which scrambles light. In the end, the MMF is imaged on the SPAD23 and creates the speckles. We measure the number of coincidences $C_{i,j}$ and single photon counts $n_i, n_j$ for each detectors $i,j$.

Then we use this information to extract the number of coincidences corresponding to real number of coincidences excluding a cross-talk and accidentals during an analogous experiment performed with SPDC. Figure \ref{Cross-talk_setup} shows the schema of an experiment for cross-talk calibration. Such a solution allows us to illuminate each detector from the SPAD23 array with different light intensities and simplify finding the cross-talk and accidental counts.
\begin{figure}[b!]
  \centering
  \includegraphics[width=1\linewidth]{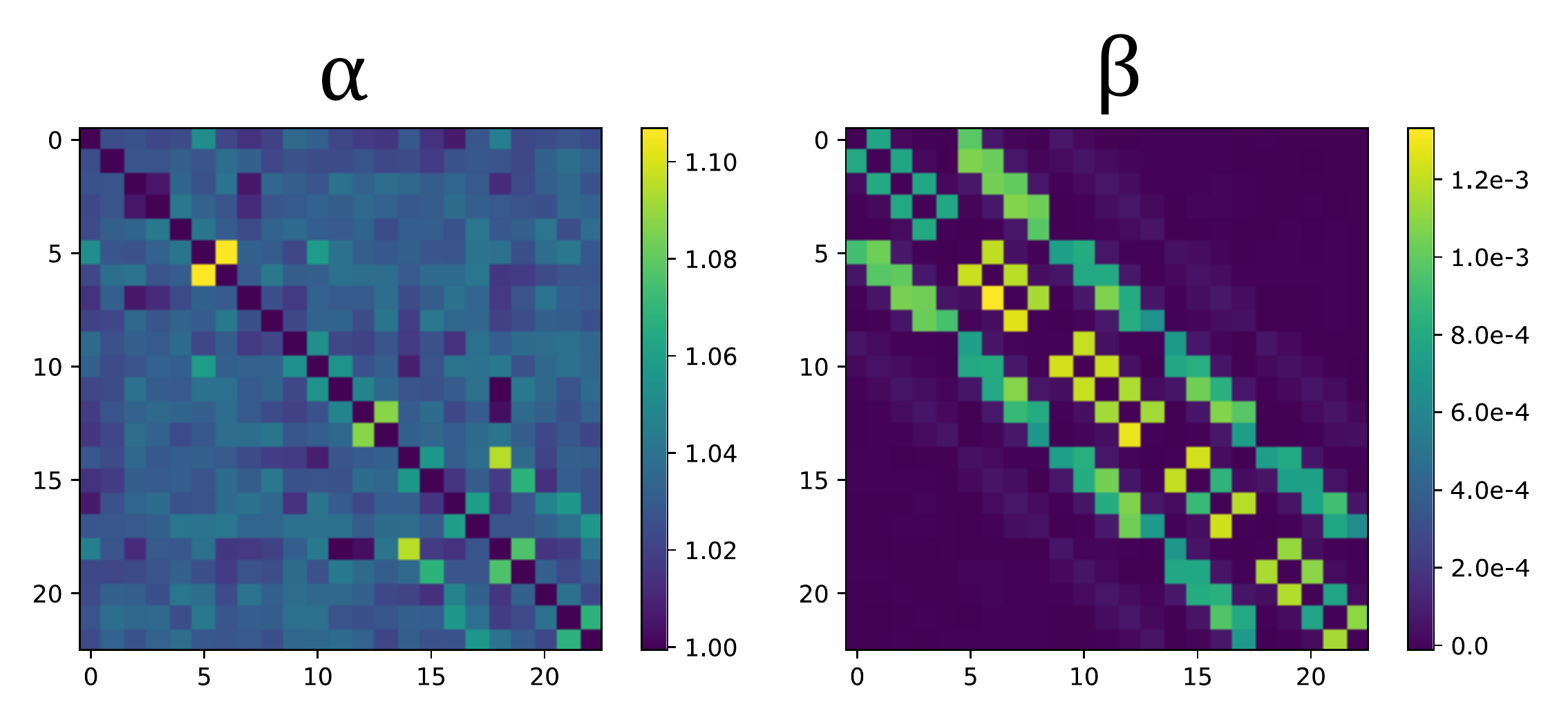}
\caption{Coefficients $\alpha$ and $\beta$ used for the accidental coincidences and cross-talk counts subtraction. The characteristic structure shown in $\beta$ matrix is defined by the SPAD23 array geometry where particular detectors are places nearer or further from each other. The presented results show that cross-talk contribution is smaller than the coincidence events comes from Poissonian distribution of incoming photon flux in the classical laser beam.}
\label{alphaandbeta}
\end{figure}
In our analysis, we assume that we measure only the coincidences caused by cross-talk and accidental counts, because of the classical nature of the light.  SPAD23 is characterized by cross-talk events linear to the number of detected photons \cite{Lubin:19}. To the data set containing $C_{i,J}$ and $n_i$, $n_j$ we fit the coefficients $\alpha_{i,j}$ and $\beta_{i,j}$ according to equation:
\begin{equation}\label{AC_final}
C_{i,j}=\alpha_{i,j}  n_i n_j \Delta t+\beta_{i,j} n_i + \beta_{j,i} n_j,
\end{equation}

where $\Delta t$ is the coincidence bin window, $\alpha_{i,j}$ and $\beta_{i,j}$  are proportional coefficients for the accidental events, and  cross-talk parameter measured by the detectors $i$ and $j$, respectively for each $s_p$ realisation on the SLM. Due to using many realisations of SLM patterns we could address various number of counts $n_i$ and coincidences $C_{i,j}$. In our calibration procedure, we create around 1000 speckle patterns to illuminate them on SPAD23. We acquire data for 1 s to measure the number of counts and coincidences for each SPAD23 detector. Thus we may smooth interpolation of $C_{i,j}$ surface for undetected number of counts.  

After obtaining matrices $\mathbf{\alpha}$ and $\mathbf{\beta}$ shown in Fig. \ref{alphaandbeta} we can estimate the number of classical light coincidences and subtract them from the total number of coincidences obtained in the SPDC-type experiment with photon pairs. Thus the HOM visibility and similarity values (described in main manuscript) can be properly calculated.

\nocite{*}
\bibliographystyle{apsrev4-1}
\bibliography{supplement} 
